\documentstyle[prl,twocolumn,aps,epsf]{revtex}

\newcommand{\extraspace}{\addtolength{\abovedisplayskip}{2mm} 
                        \addtolength{\belowdisplayskip}{2mm} 
                        \addtolength{\abovedisplayshortskip}{2mm} 
                        \addtolength{\belowdisplayshortskip}{2mm}} 
\newcommand{\be}{\begin{equation}\extraspace} 
\newcommand{\ee}{\end{equation}} 
\newcommand{\bea}{\begin{eqnarray}\extraspace} 
\newcommand{\eea}{\end{eqnarray}} 
 
\newcommand{\nonu}{\nonumber \\[2mm]} 
 
\newcommand{\strutje}{\rule[-1mm]{0mm}{4mm}} 
\newcommand{\strutj}{\rule[-2mm]{0mm}{4mm}}

\newcommand{\half}{{\textstyle \frac{1}{2}}} 
\newcommand{\quart}{\frac{1}{4}} 
\newcommand{\threequart}{\frac{3}{4}}

\newcommand{\vac}{| 0 \rangle}

\newcommand{\tr}{{\strutje \rm trace \,}}

\newcommand{\del}{\partial} 
\newcommand{\eg}{{\it e.g.}} 
\newcommand{\ie}{{\it i.e.}} 
\newcommand{\cW}{{\cal W}}

\begin{document} 
 
\title{Exclusion Statistics in Conformal Field Theory Spectra} 
\author{Kareljan Schoutens} 
\address{ 
     Institute for Theoretical Physics,
     Valckenierstraat 65, 1018 XE Amsterdam, 
     THE NETHERLANDS} 
\date{June 16, 1997}
\maketitle
\begin{abstract} 
We propose a new method for investigating the exclusion  
statistics of quasi-particles in Conformal Field Theory  
(CFT) spectra. The method leads to one-particle distribution
functions, which generalize the Fermi-Dirac distribution.
For the simplest $su(n)$ invariant CFTs we find a generalization 
of Gentile parafermions, and we 
obtain new distributions for the simplest $Z_N$-invariant 
CFTs. In special examples, our approach reproduces distributions  
based on `fractional exclusion statistics' in the sense of 
Haldane. We comment on applications to fractional quantum
Hall effect edge theories. 
\end{abstract}
\vskip 2mm
\noindent{\small Report number: ITFA-97-19}

\narrowtext
 
\section{Introduction} 
 

Conformal Field Theory (CFT) in two dimensions is an
invaluable tool in the analysis of (among other things) 
the low-temperature properties of a variety of Condensed 
Matter systems. In the 
literature on CFT (which is vast), there is a certain 
dichotomy between, on the one hand, descriptions based on 
{\em bosonization}\ 
and, on the other, descriptions which give a central role to 
{\em quasi-particles}.
 

In the standard approach to rational CFT, the spectrum is 
described in terms of representations of a
bosonic current algebra called the {\em chiral algebra}. 
Examples are affine Kac-Moody (KM) algebras and higher spin 
extensions (called $\cW$-algebras) of the Virasoro algebra. 
In applications to Condensed Matter systems, a similar
description is often used. Examples are the Luttinger 
Liquids for 1D interacting electrons, which have a $U(1)$ affine 
KM symmetry and are usually treated in bosonized form. 
Other examples are the low-temperature theories 
for the multi-channel Kondo effect, which have been analyzed 
on the basis of their $su(2)_k$ affine KM symmetry.


When dealing with the CFT for non-interacting electrons, one
clearly does not need bosonization, but uses free fermions 
(satisfying canonical anti-commutation relations) instead. While 
interacting electrons give rise to more general CFTs, 
it is entirely natural to look for descriptions that mimic 
the treatment of free electrons. The idea is to identify 
fundamental excitations (quasi-particles) over the many-body 
ground state and to study their properties. For integrable models 
(analyzed using Bethe Ansatz and factorizable scattering) 
such an approach is by now standard. 
 
Until now, a general approach to `CFT quasi-particles' has been lacking. 
In special cases, progress has been made by viewing specific CFTs 
as massless limits of integrable particle theories, leading to 
`massless S-matrices for CFT quasi-particles'. Related to this
is a new `integrable' approach to CFT, see \cite{BLZ}. Other special 
examples are CFTs which can be viewed as continuum limits of 
integrable models of lattice electrons. Examples are the $su(n)_1$ 
CFTs, which can be cast in a spinon formulation analogous to that 
of the $su(n)$ Haldane-Shastry (HS) spin chains \cite{BS}. 


In this Letter, we propose an approach to CFT 
quasi-particles which is intrinsic to the CFT, \ie, which does not 
make reference to associated integrable particle or lattice models. 
The starting point is the finite size spectrum of a CFT 
defined on a cylinder. In particular, we focus on the chiral 
Hilbert spaces, which together build up a CFT partition function.  
In a quasi-particle formulation, a chiral Hilbert space is viewed 
as a collection of multi-(quasi-)particle states%
\footnote{We shall write `particle' for `quasi-particle'
          where no confusion can arise.}. 
For {\em fermionic}\ quasi-particles, the systematics of 
multi-particle states are simply given by the Pauli Principle,
resulting in the Fermi-Dirac distribution function. For more 
general quasi-particles one 
may try to understand the spectrum in terms of more general 
distributions that correspond to various forms of 
{\em exclusion statistics}. 
 
The notion of exclusion statistics was introduced by Haldane \cite{Ha1}, 
in the context of integrable theories with 
inverse square interactions.  
The main idea to study the way one-particle levels in the 
spectrum are filled to form allowed many-particle states. The 
simplest scenario \cite{Ha1} is to 
assume that the act of filling an available one-particle state of type 
$i$ reduces the dimension of the available Hilbert space for particles 
of type $j$ by an amount $g_{ij}$. In the absence of other interactions,
the statistics matrix $G=(g_{ij})$ completely determines the 
thermodynamics (see \eg \cite{gstats,IAMP}). Concrete 
examples of this type of exclusion statistics are the 
Calogero-Sutherland (CS) models of quantum mechanics with inverse 
square two-body interactions (with adjustable $g$).
 

We here propose a new method for studying the 
exclusion statistics of CFT quasi-particles. At the heart of our 
method is what can be called  a `transfer matrix for truncated 
chiral spectra'.  
We shall present a number of examples where CFT spectra are 
completely encoded in one-particle 
distribution functions (generalizing the Fermi-Dirac distribution). 
In special examples, the statistics that we find are
of the type proposed by Haldane, while in other cases we find 
more general results. 
 
One check on the distributions that we propose here is the coefficient 
$\gamma$ in the low temperature specific heat, which is known to be 
related to the central charge $c_{CFT}$ of the CFT according 
to \cite{cCFT} 
\be
  {C \over L} = \gamma k_B^2 \rho_0 \, T \ , \qquad 
  \gamma = {\pi \over 6} \, c_{CFT} \ ,
\ee
where $\rho_0=(\hbar v_F)^{-1}$ is the density of states
per unit length. 

\section{Introducing the method}

To introduce our new method we focus on the simplest $su(2)$
invariant CFT, which is the $su(2)_1$ Wess-Zumino-Witten
model. For this CFT, a quasi-particle formulation has been 
proposed in \cite{su2spi} and worked out in great detail in 
\cite{su2cft}. The formulation uses
operators $\phi^{\pm}_{-s}$ which create quasi-particles called
spinons. The spinons form a doublet under $su(2)$ and carry 
(dimensionless) energy $L_0=s$. The chiral spectrum of the
$su(2)_1$ CFT may be built in the following manner. One starts by 
writing the following {\em polarized $N$-spinon states}
\be
\phi^+_{-{2N-1 \over 4}-n_N} \ldots
\phi^+_{-{3\over 4}-n_2} \phi^+_{-{1\over 4}-n_1} \vac
\ee
with $n_N \geq \ldots \geq n_2 \geq n_1 \geq 0$.
One then uses the Yangian symmetry algebra to construct
multi-spinon states with mixed $+$ and $-$ indices. The 
collection
of all these states forms a basis of the full chiral spectrum. 
The $su(2)$ content of the Yangian multiplet labeled by 
$n_1, \ldots n_N$ follows from the generalized commutation relations 
satisfied by the spinon modes, or, equivalently, from the 
representation theory of the Yangian \cite{su2spi,su2cft}.

Comparing the allowed spinon modes $\phi^\pm_{-s}$
with free fermion modes $\psi_{-\half-l}$, we observe that 
the fermion mode $l+\half$ has `split' into an odd mode 
$s=l+\quart$ and an even mode $s=l+\threequart$. To maintain a 
spacing of one unit, we view these 
two spinon modes as forming a single `one-particle level' 
in the spectrum. What we would like to do, is to factorize the 
full chiral partition sum of the CFT into a product over 
these one-particle levels, so that the free energy becomes 
a sum of one-particle contributions. 

While we can not straightforwardly extract the contribution
of the $l^{\rm th}$ level, we may proceed as follows.
We truncate the chiral spectrum by only allowing the spinon 
modes of the first $l$ levels. 
We denote by $P_l(q,x_+,x_-)$, $Q_l(q,x_+,x_-)$ the partition 
functions of the first $l$ levels,
where in $P_l$ the highest occupied mode should be odd
while in $Q_l$ it should be even. In formula
\be
P_l(q,x_+,x_-) = \tr^{(\leq l)}_{\rm odd} 
  ( q^{L_0} x_+^{N_+} x_-^{N_-}) 
\ee
and similar for $Q_l(q,x_+,x_-)$, where we introduce chemical 
potentials $\mu_{\pm}$, write 
$x_{\pm}=e^{\beta\mu_{\pm}}$, 
$q=e^{-\beta {2\pi \over L}{1 \over \rho_0}}$, and denote 
by $N_{\pm}$ the number of $\pm$ spinons in a state. 

Restricting to fully polarized states (putting $x_-=0$), we
obtain the following recursion relation 
\be
\left( \begin{array}{c} P_l \\ Q_l \end{array} \right)
=
\left( \begin{array}{cc} 
                  \strutje 1  & 
                  q^{l-\threequart} \, x_+  \\
                  q^{l-\quart} \, x_+  & 
                  1 + q^{2l-1} \, x_+^2 
       \end{array} \right)
\left( \begin{array}{c} P_{l-1} \\ Q_{l-1} \end{array} \right)
\label{recM1}
\ee
with $P_0=0$ and $Q_0=1$. We now make the important step
of approximating these exact expressions for $P_l(q,x_+)$
and $Q_l(q,x_+)$ by the product from $j=1$ to $l$ of the 
largest eigenvalue $\lambda_+^{(j)}(q,x)$ of the $j^{\rm th}$ 
recursion matrix. Clearly, this brings
the partition sums in the desired factorized form and
reduces thermodynamic quantities to sums of independent 
single-level contributions! The average occupation of
the $l^{\rm th}$ level is found to be
\bea
&& \overline{n}^{(l)}(q,x)  
   = x \partial_x \log \lambda_+^{(l)}(q,x) 
   = {2  \over \sqrt{\strutje 1+ 4 \, q^{1-2l}x^{-2}}} 
\nonu
   && \qquad 
   = {2 \over \sqrt{\strutje 1+4 \, 
     e^{-2 \beta (\mu_+-\epsilon_l)}}} \ .
\eea
with $\epsilon_l = (l+\half){2\pi \over L}{1 \over \rho_0}$.
Note that the maximal occupation of a given level is 2. 
Interestingly, this distribution
function for polarized $su(2)$ spinons is identical to
that of particles with $g=\half$ (semionic) exclusion
statistics \cite{gstats}.

If we now include the multi-spinon states with mixed
indices, we find that the $l^{\rm th}$ recursion matrix 
becomes
\be
\left( \! \begin{array}{cc} 
           \strutj 1-q^{2l-2}x^2 &  
           q^{l-\threequart} x (z+{1 \over z}) \\
           q^{l-\quart} x (z+{1 \over z}) (1-q^{2l-2}x^2) & 
           1 + q^{2l-1} x^2 (z^2 + 1 + {1 \over z^2})
       \end{array} \! \right)
\label{recM2}
\ee
where we put $x_{\pm}=x \, z^{\pm 1}$.
Following the same logic, we derive the distribution functions   
$\overline{n}^{(l)}(q,x,z)$. Interesting special case are
\be
\overline{n}^{(l)}(q,x)  
   = x \partial_x \log \lambda_+^{(l)}(q,x,z=1) 
   =  {2 \over 1+q^{\half-l}x^{-1}}
\label{disx}
\ee
for the expected total number of spinons in level $l$, 
and
\bea
&& \overline{Q}^{(l)}(q,z)  
   = {e \over 2} \, z \partial_z \log \lambda_+^{(l)}(q,x=1,z) 
\nonu   
&& = {e \, q^{l-\half} (z-z^{-1}) \over
      \strutje \sqrt{ \strutje q^{2l-1} (z+z^{-1})^2 + 4 \, (1-q^{2l-1})} }
\label{disz}
\eea
for the expected charge at level 
$l$ (we assume charges $\pm {e \over 2}$ for the $\pm$ spinons). 
The distributions 
(\ref{disx}) and (\ref{disz}) agree with the distributions 
obtained from fractional exclusion statistics
with $G = \left( \begin{array}{cc} \strutj \half & \half \\ 
\half & \half \end{array} \right)$. Note that
(\ref{disx}) implies that for $z=1$ (zero voltage)
the thermodynamics of the spinon system is identical to that 
of two free fermions and the central charge is 
$c_{CFT}=2 \times \half =1$.
Note also that, with $z=e^{\half \beta e V}$, the integrated 
charge ${1 \over L}\sum_{l=1}^\infty  \overline{Q}^{(l)}(q,z)=
{1 \over 4\pi} e^2 V \rho_o$, which is half of the value obtained for
two charge $\pm e$ free fermions.

The correspondence with Haldane statistics is satisfying 
since the spinons of the associated 
$su(2)$ HS spin chain satisfy these same statistics \cite{Ha1}. 
This confirms the validity of our new approach, which in no 
way relied on the exact solution of the HS chain. 

\section{Examples}

\subsection{The $su(n)_1$ CFTs}

The first generalization of the $su(2)_1$ results concerns
the $su(n)_1$ spinons. The yangian symmetry of the $su(n)_1$
CFT was established in \cite{Sc} while the spinon formulation 
was presented in \cite{BS}. There are $n$ fundamental spinon
species $\phi^i$, 
transforming in the representation $\overline{\bf n}$ of $su(n)$.
Repeating the analysis shown above, finding explicitly the
$su(n)$ analogue of the recursion matrices 
(\ref{recM1}) and (\ref{recM2}) (see \cite{Sc2}), we find that
\begin{enumerate}
\item
a single spinon species $\phi^{i_0}$ (in absence of any others) 
satisfies Haldane statistics with $g={n-1 \over n}$, 
\item
when exciting all $n$ spinon species symmetrically
(choosing all $x_i$ equal to $x=e^{\beta\mu}$), the
expected total occupation of the $l^{\rm th}$ level
is given by
\be
  \overline{n}^{(l)}(q,x) =  
  x \partial_x \log[1+q^l x+ \ldots+ (q^l x)^{n-1}]^n.
\label{gentile}
\ee
\end{enumerate}
Comparing these results, one finds that for $n>2$ 
there is negative mutual exclusion among different spinons.

Interestingly, the  statistics going with the distribution
(\ref{gentile}) were proposed by Gentile as early 
as 1940 \cite{Ge}. One finds that a single `Gentile parafermion' 
contributes the amount ${n-1 \over n}$ to the central charge, 
so that the full result becomes
$c_{CFT} = n {n-1 \over n} = n-1$. We do not expect
that a single Gentile parafermion with $n>2$ can define a 
consistent CFT spectrum. For example, the case $n=3$ would 
lead to a $c_{CFT}={2 \over 3}$ CFT, for which good (unitary) 
candidates are lacking.

Our results are consistent with \cite{KK}, where, by 
different methods, the link between $su(n)$ HS spin chains 
and Gentile parafermions has also been established. 

\subsection{$Z_N$ parafermions}

Within the context of CFT, the simplest generalization of the 
Majorana fermion is the so-called $Z_N$ parafermion.
It features in a CFT of central charge $c_N={2(N-1) \over N+2}$
as a primary field of dimension $h_N={N-1 \over N}$
and $Z_N$ charge 1 \cite{FZ}. By applying the method outlined 
above, we obtained a distribution function 
for the $Z_N$ parafermion and established that the full CFT spectrum
is reproduced by a gas of non-interacting quasi-particles of this 
type. For the purpose of explaining these results, we focus
on the case $N=3$.

It is well known that the chiral spectrum of the $Z_3$ parafermion
CFT can be interpreted in terms of two parafermions $\psi^{\pm}$
of opposite $Z_3$ charge. 
However, by exploiting the generalized commutation relations obeyed
by the modes of $\psi^{\pm}$ \cite{FZ}, one easily shows that the
modes of $\psi^+$ alone can generate the full spectrum, and that
the $\psi^-$ quasi-particle can be viewed as a composite of two 
$\psi^+$ quasi-particles. Having understood how the $\psi^+$ modes 
alone build the chiral spectrum (see \cite{Sc2}), one may define 
truncated partition sums.
We found the following recursion matrix between the $l^{\rm th}$ 
and the $(l-1)^{\rm th}$ truncated sums (which each have 
three components)
\be
\left( \begin{array}{ccc} 
           (1-y^3)     & y^2      &  y     \\
           y(1-y^3)    &  1       & 2y^2   \\
           2y^2(1-y^3) & y(1+y^3) & 1+2y^3
       \end{array} \right) ,
\label{recZ3}
\ee
with $y=x \, q^l$. The $Z_3$ parafermion distribution function is 
expressed in terms of the largest eigenvalue $\lambda_+$ 
of this matrix
\be
\overline{n}^{(l)}(q,x) 
   = (y \del_y \log \lambda_+)(y= x \, q^l) \ .
\ee
From the asymptotics $\lambda_+(y)\propto y^3$, one finds that 
the maximum occupation per level equals 3. 
Using the Cardano formula one may write $\overline{n}^{(l)}$ 
in closed form. We here present a plot (Fig.~1), which 
displays the function $\overline{n}$ as a function of energy.
As a check we (numerically) evaluated the coefficient 
$\gamma$ of the specific heat, reproducing the expected 
value $\gamma = {\pi \over 6}\, {4 \over 5}$. 

For general $N$, the distribution for $Z_N$ parafermions 
allows a maximum of $\half N (N-1)$ particles per level.

\subsection{Quantum Hall Effect edge theories}

As a further application we briefly discuss edge theories for 
the Fractional Quantum Hall Effect (FQHE). We shall come back to 
this topic in a separate publication \cite{ES}. For the 
$\nu={1 \over m}$ 
FQHE (with $m$ an odd integer), the edge theory is a chiral $c_{CFT}=1$ 
CFT at compactification radius $R^2=m$. The natural quasi-particles
to consider are the edge electron (of charge $-e$) and the edge 
quasi-hole (of charge ${e \over m}$). Writing the spectrum in terms of 
these quasi-particles, and applying the above procedure, we find that
the fundamental quasi-particles are independent, and obey
Haldane exclusion statistics with $g=m$ and $g=1/m$, respectively. 
This is consistent with the result of bosonization applied 
to $g$-ons \cite{WY}.

The (known) results for the specific heat ($c_{CFT}=1$)
and the response to voltage (${Q \over L} = {1 \over m} 
{1 \over 2\pi} e^2 V \rho_0 $)
are easily reproduced by exploiting the duality between 
$g=m$ and $g=1/m$ statistics \cite{ES}. The central charge
arises as a sum $c_{m}+c_{1/m}$. For $m=2$ we find 
$c_{2}={2 \over 5}$,  $c_{1/2}={3 \over 5}$, while for $m=3$,
$c_{3}={0.343\ldots}$,  $c_{1/3}={0.655\ldots}$.
  
These quantum Hall results can be appreciated on the basis of 
the analogy with CS quantum mechanics with inverse
square interactions (see \eg \cite{Is}). We stress, 
however, that our derivation does not rely on this analogy.

When applied to composite edges for the FQHE in the Jain series,
at filling fraction $\nu = {n \over 2np+1}$, the combined results
of this paper lead to a formulation in terms of 
(i) a single charged mode, satisfying Haldane statistics with 
$g=\nu$, 
and (ii) a set of $n$ spinons for $su(n)_1$, satisfying the
generalized Gentile statistics described above. This new 
quasi-particle formulation forms a suitable starting point for 
studying finite-$T$ features (including tunneling characteristics) 
of these edges.

\section{Applications and outlook}

The potential applications of our new approach to CFT spectra 
are manifold, especially when the extension to boundary CFTs is 
considered. We mention edge state scattering, state counting
and a variety of finite-$T$ characteristics of (non-abelian) FQHE
edges, and non-Fermi liquid features in quantum impurity problems 
such as the multi-channel Kondo effect.

The method presented here can successfully be applied
to many CFTs other than those mentioned here. An interesting 
example is the $c_{CFT}=-{22 \over 5}$ CFT for the Yang-Lee 
edge singularity where the Virasoro generators take on the role 
of $g=2$ quasi-particles, giving the correct effective central
charge $\tilde{c}_{CFT} = {2 \over 5}$ \cite{FS}.

We should stress that the distribution functions presented 
here are different from the distributions obtained from
massless $S$-matrices using the Thermodynamic Bethe Ansatz. 
The relation between the two approaches is presently not clear.

\vskip 3mm

\noindent {\bf Acknowledgements.}\  
Many thanks to A.W.W.~Ludwig, P.~Bouw\-knegt
and R.~van Elburg for helpful discussions. Part of this work
was done at the ITP Santa Barbara Workshop on `Quantum Field 
Theory in Low Dimensions: from Condensed Matter to Particle Physics'.
The author is supported in part by the foundation FOM.

\vbox{
\epsfxsize=9.5cm
\epsfysize=11cm
\epsffile{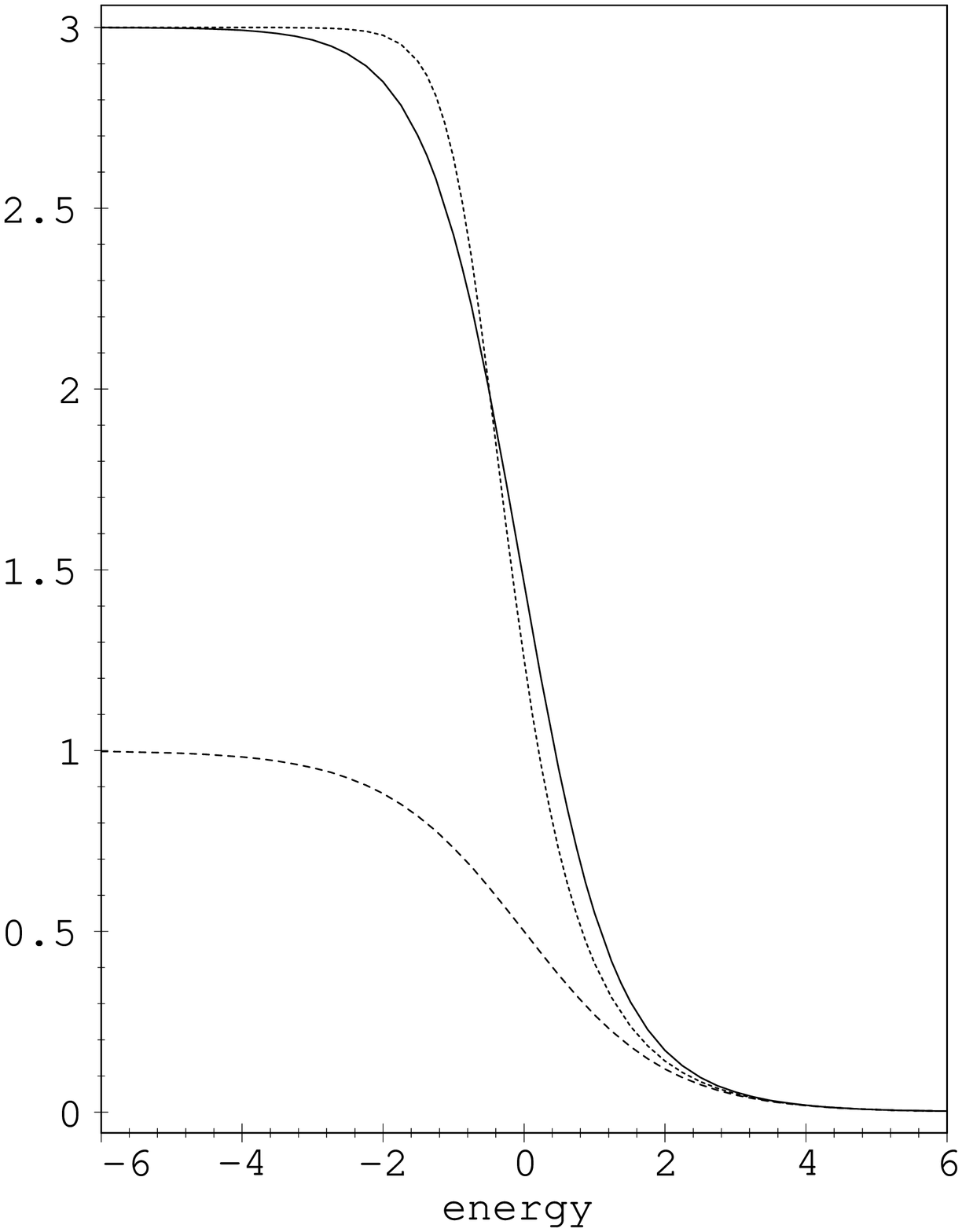}
\begin{figure}
\caption[]{Distribution functions 
  for $Z_3$ parafermions (solid line), for ordinary
  fermions (dashed line), and for particles satisfying 
  $g={1 \over 3}$ exclusion statistics (dotted line).
  All distributions are at $\mu=0$; the energy is given 
  in units $\beta^{-1}$.}
\end{figure}
}
   

\begin{references} 

\bibitem{BLZ}
  V. Bazhanov, S. Lukyanov and A.B. Zamolodchikov,
  Comm. Math. Phys. {\bf 11}, 381 (1996);
  hep-th/9604044.
\bibitem{BS} 
  P. Bouwknegt and K. Schoutens, Nucl. Phys.  
  {\bf B482}, 345 (1996); q-alg/9703021.
\bibitem{Ha1} 
  F.D.M. Haldane, Phys. Rev. Lett. {\bf 67}, 937 (1991). 
\bibitem{gstats}
  Y.S. Wu, Phys. Rev. Lett. {\bf 73}, 922 (1994);
  C. Nayak and F. Wilczek, Phys. Rev. Lett {\bf 73}, 2740 (1994);
  S.B. Isakov, Mod. Phys. Lett. {\bf B8}, 319 (1994). 
\bibitem{IAMP}
  S.B. Isakov, D.P. Arovas, J. Myrheim, A.P. Polychro\-nakos,
  Phys. Lett. {\bf A212}, 299 (1996).
\bibitem{cCFT}
  H.W.J. Bl\"ote, J.L.  Cardy and M.P. Nightingale,
  Phys. Rev. Lett. {\bf 56}, 742 (1986);
  I. Affleck, Phys. Rev. Lett. {\bf 56}, 742 (1986).
\bibitem{su2spi} 
  F.D.M. Haldane, Phys. Rev. Lett. {\bf 66}, 1529 (1991); 
  F.D.M. Haldane, Z.N.C. Ha, J.C. Talstra, D. Bernard 
  and V. Pasquier, Phys. Rev. Lett. {\bf 69}, 2021 (1992). 
\bibitem{su2cft} 
  D. Bernard, V. Pasquier and D. Serban, 
  Nucl. Phys. {\bf B428}, 612 (1994);
  P.~Bouwknegt, A.W.W. Ludwig and K. Schoutens, 
  Phys. Lett. {\bf B338}, 448 (1994).
\bibitem{Sc}
  K. Schoutens, Phys. Lett. {\bf B331}, 335 (1994).
\bibitem{Sc2}
  K. Schoutens, in preparation.
\bibitem{Ge}
  G. Gentile, Nuovo Cimento {\bf 17}, 493 (1940).
\bibitem{KK}
  Y. Kato and Y. Kuramoto, J. Phys. Soc. Jap.
  {\bf 64}, 4518 (1995).
\bibitem{FZ}
  A.B. Zamolodchikov and V.A. Fateev, Sov. Phys.
  JETP {\bf 62}, 215 (1985).
\bibitem{ES}
  R. van Elburg and K. Schoutens, in preparation.
\bibitem{WY}
  Y.-S. Wu and Y. Yu, Phys. Rev. Lett. {\bf 75},
  890 (1995).
\bibitem{Is}
  S. Iso, Nucl. Phys. {\bf B443}, 581 (1995).
\bibitem{FS}
  E. Frenkel and A. Szenes, Duke Math. J.
  {\bf 69}, 53 (1993).
 
\end{references}
\end{document}